\documentclass[%
reprint,
superscriptaddress,
aps,
pre,
]{revtex4-2}
\usepackage{graphicx}
\usepackage{dcolumn}
\usepackage{bm}
\usepackage{amsfonts}
\usepackage{float}
\usepackage{amsmath}
\usepackage{amssymb}
\usepackage{comment}
\usepackage{acronym}
\usepackage{wasysym}

\usepackage{bbding}

\usepackage{color}

\newacro{PE}{participation entropy}
\newacro{PC}{participation coefficient}

\newcommand{\fig}[1]{Fig.~\ref{fig:#1}}
\newcommand{\eq}[1]{Eq.~(\ref{eq:#1})}

\begin{document}


\title{On the information-theoretic formulation of network participation}

    \author{Pavle Cajic}
    \affiliation{School of Physics, Faculty of Science, The University of Sydney, Sydney NSW 2006, Australia}

    \author{Dominic Agius}
    \affiliation{School of Physics, Faculty of Science, The University of Sydney, Sydney NSW 2006, Australia}

    \author{Oliver M. Cliff}
    \affiliation{School of Physics, Faculty of Science, The University of Sydney, Sydney NSW 2006, Australia}
    \affiliation{Centre for Complex Systems, The University of Sydney, Sydney NSW 2006, Australia}

    \author{James M. Shine}
    \affiliation{Centre for Complex Systems, The University of Sydney, Sydney NSW 2006, Australia}
    \affiliation{Brain and Mind Centre, Faculty of Medicine, The University of Sydney, Sydney NSW 2006, Australia}

    \author{Joseph T. Lizier$^\ast$}
    \affiliation{Centre for Complex Systems, The University of Sydney, Sydney NSW 2006, Australia}
    \affiliation{School of Computer Science, Faculty of Engineering, The University of Sydney, Sydney NSW 2006, Australia}

    \author{Ben D. Fulcher$^\ast$}
    \affiliation{School of Physics, Faculty of Science, The University of Sydney, Sydney NSW 2006, Australia}
    \affiliation{Centre for Complex Systems, The University of Sydney, Sydney NSW 2006, Australia}

%
%
%


\begin{abstract}
The participation coefficient is a widely used metric of the diversity of a node's connections with respect to a modular partition of a network.
An information-theoretic formulation of this concept of connection diversity, referred to here as participation entropy, has been introduced as the Shannon entropy of the distribution of module labels across a node's connected neighbors.
While diversity metrics have been studied theoretically in other literatures, including to index species diversity in ecology, many of these results have not previously been applied to networks.
Here we show that the participation coefficient is a first-order approximation to participation entropy and use the desirable additive properties of entropy to develop new metrics of connection diversity with respect to multiple labelings of nodes in a network, as joint and conditional participation entropies.
The information-theoretic formalism developed here allows new and more subtle types of nodal connection patterns in complex networks to be studied.
\end{abstract}

\maketitle


Many real-world networks exhibit modular structure, in which nodes form densely interconnected modules with relatively sparse connectivity between modules.
Such modularity is observed in social networks, food webs, metabolic networks, protein--protein interaction networks, air-traffic networks, and brain networks \cite{guimera2005cartography}.
Within such modular networks, individual nodes can vary substantially in their degree of within- versus across-module connectivity.
These differences can provide important insights into a node's functional role within a network, such as facilitating local information processing (consistent with strong within-module connectivity) versus distributed/integrative communication (strong cross-module connectivity).

To measure the extent to which a given node's connections are distributed within or across modules, the participation coefficient was introduced by Guimer\`a and Amaral~\cite{Guimera2005, guimera2005cartography}.
It has been used widely to analyze networks across domains, including the Internet, metabolic, air transportation, protein-interaction, and neural networks \cite{guimera2007classes, Sporns2007}.
For example, the participation coefficient of nodes in macroscopic brain networks has been used to distinguish levels of consciousness caused by brain injury \cite{Rizkallah2019:DecreasedIntegrationEEG} and to identify emerging new research directions from scientific publication citation networks \cite{Shibata2008:DetectingEmergingResearch}.
This concept of nodal connection diversity across modules was also formulated as a Shannon entropy by \citet{RubinovSporns}.
Quantifying diversity is a general problem studied across many fields, with a prominent application to species diversity in ecology for which the Shannon entropy and Gini--Simpson index (the measure underlying the participation coefficient \cite{RubinovSporns}) formulations have been used for decades, among a host of alternative indices \cite{Peet2003, Daly2018}.
Mathematical relationships between different formulations of diversity indices have been uncovered.
For example, the Gini--Simpson index and Shannon entropy have each been shown to be special cases of `generalised entropies' \cite{Havrda1967, Keylock2005, Vajda2007}.
\citet{Zhang2014} have further shown that the Gini--Simpson index can be expressed as a first-order Taylor approximation to the Shannon entropy formulation of diversity.

Despite the wide variety of diversity indices used in ecology, the participation coefficient has remained the dominant measure of node participation in network theory since it was introduced in 2005 \cite{Guimera2005, guimera2005cartography}.
Here we connect the problem of quantifying nodal connection diversity in networks with a large and existing literature on diversity indices, and in particular explain the relationship between the participation coefficient and the corresponding Shannon entropy-based formulation of connection diversity \cite{RubinovSporns}, which we call `participation entropy' here.
We argue that participation entropy is a better-motivated measure of node participation diversity, primarily due to its additive behaviour with respect to chaining probability distributions, an operation which can arise naturally when the nodes in a network have labels in multiple module sets.
Taking advantage of this behavior, we define novel measures of connection diversity---`joint' and `conditional' participation entropy---for quantifying more nuanced types of connection patterns in complex networks.

\subsection*{Participation Coefficient and Participation Entropy}

We consider a binary, undirected network partitioned into $M$ non-overlapping modules, with each node labeled as belonging to a module, from the set $\mathcal{M} = \{m_1,m_2,...,m_M\}$.
Note that this modular partition is most commonly obtained as the result of a community-detection algorithm operating on the network \cite{fortunato2016community}, but could in general represent any assignment of categorical labels to nodes in a network.
Given $\mathcal{M}$, the participation coefficient, $\mathcal{P}_i$, of node $i$ is defined as
\begin{equation}\label{eqn:PC}
    \mathcal{P}_i(\mathcal{M}) = 1 - \sum_{j = 1}^{M} \left( \frac{\kappa_{ij}}{k_i} \right)^2 \,,
\end{equation}
where $\kappa_{ij}$ is the number of edges between node $i$ and a node in module $m_j$, and $k_i$ is the degree of node $i$ (the total number of connections made to all other nodes in the network) \cite{Guimera2005, guimera2005cartography}.
For simplicity, we focus on undirected networks here, but note that this formulation extends straightforwardly to weighted networks (substituting $\kappa_{ij}$ and $k_i$ for weighted versions that sum edge weights) and directed networks (e.g., by defining $\kappa_{ij}$ and $k_i$ as counting connections outward from, or arriving to node $i$, as the in-degree or out-degree).
Equation~\eqref{eqn:PC} exhibits the desired behavior of a connection diversity metric, taking a minimal value for a node with connections entirely within a single module ($\mathcal{P}_i = 0$) and a maximal value for a node that connects equally across all $M$ modules ($\mathcal{P}_i = 1 - 1/M$).
%

\begin{figure}
    \centering
    \includegraphics[width=.9\columnwidth]{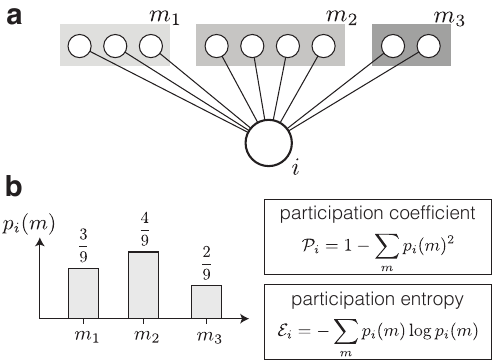}
    \caption{
    \textbf{A probabilistic formulation of a node's connection diversity with respect to a set of labels.}
    \textbf{a} We plot the connected neighbors of a given node, $i$, which span three labeled modules: $m_1$ (3 edges), $m_2$ (4 edges), and $m_3$ (2 edges).
    \textbf{b} This pattern can be represented as a probability distribution, $p_{i}(m)$, that captures the probability of node $i$'s connected neighbors being in each of the modules.
    The participation coefficient, $\mathcal{P}_i$, and participation entropy, are then computed from $p_{i}(m)$.
    }
    \label{fig:schematic}
\end{figure}

\paragraph{A probabilistic formulation.}
An alternative interpretation of Eq.~\eqref{eqn:PC} can be considered by identifying $\kappa_{ij}/k_i$ as the probability, $p_i(m_j)$, that a randomly selected connected neighbor of node $i$ is assigned to module $m_j$.
An example is depicted in Fig.~\ref{fig:schematic}a, which depicts the connected neighbors of node $i$ across each of three modules, $\mathcal{M} = \{m_1, m_2, m_3\}$.
This, or any other pattern of connectivity, can be represented as a probability distribution, $\{p_i(m)\}_{m\in\mathcal{M}}$, plotted for this simple example in Fig.~\ref{fig:schematic}b.
In this probabilistic formulation, $\mathcal{P}_i$ can be expressed as a function of $p_i(m)$ by rewriting Eq.~\eqref{eqn:PC} as $\mathcal{P}_i(\mathcal{M}) = 1 - \sum_{m \in \mathcal{M}} p_{i}(m)^2$.

This formulation allows us to clearly see that the participation coefficient is an implementation of the Gini--Simpson index of diversity \cite{Simpson1949:MeasurementDiversity}, as observed previously \cite{RubinovSporns}.
This is an important measure used in many other contexts, including quantifying biodiversity \cite{Peet2003, Daly2018}.
Following the interpretation that motivated Simpson's original formulation \cite{Simpson1949:MeasurementDiversity}, $\mathcal{P}_i$ can be interpreted as the probability that two randomly selected nodes connected to node $i$ (with replacement) lie in different modules.


\paragraph{Participation Entropy}

The Shannon entropy \cite{Shannon1948, Cover2005} of $p_{i}(m)$ is a natural measure of the connection diversity of node $i$ across the label set, $\mathcal{M}$:
\begin{equation} \label{eq:participation-entropy}
    \mathcal{E}_i(\mathcal{M}) = H[p_i(m)] = -\sum_{m \in \mathcal{M}} p_{i}(m) \log p_{i}(m)\,.
\end{equation}
We term $\mathcal{E}_i(\mathcal{M})$ the `participation entropy' of node $i$, which measures the uncertainty (or average surprise) in the module labels (from $\mathcal{M}$) of its connected neighbors.
This matches a previous formulation of nodal connection diversity introduced by \citet{RubinovSporns} (named the `diversity coefficient' in its implementation in code in the \textit{Brain Connectivity Toolbox} \cite{Rubinov2010:ComplexNetworkMeasures}).
Participation entropy exhibits the same desired qualitative behavior as the participation coefficient, $\mathcal{P}_i$; that is, $\mathcal{E}_i = 0$ is minimal when all connected neighbors of node $i$ are in the same module (minimum uncertainty about the module label of node $i$'s neighbors) and $\mathcal{E}_i = \log(M)$ is maximal when connected neighbors are equally distributed across all of the modules (maximum uncertainty about the module label of node $i$'s neighbors).
Note that both $\mathcal{E}_i$ and $\mathcal{P}_i$ may be normalised by dividing by their maximum value for a given number of modules $M$, if desired (as `normalized connection diversity' \cite{RubinovSporns}, which has the effect of setting its range to the unit interval).

Compared to $\mathcal{P}_i$, quantifying connection diversity as an entropy, $\mathcal{E}_i$, is the unique formulation that satisfies three key advantageous properties.
First, it is continuous with respect to changes in $p_{i}(m)$.
Second, it increases monotonically with the number of modules, $M$, when $p_{i}(m) = 1/M$, $\forall m$.
Third, and most importantly, $\mathcal{E}_i$ can be chained consistently across multiple labeling sets for nodes \cite{Shannon1948, Ash1965}, opening new ways of quantifying and interpreting nodal connection patterns in networks, as we develop later.
As per the original formulation of $\mathcal{P}_i$, it also generalizes straightforwardly to weighted and directed networks.

\paragraph{Connecting the two formulations}

The mathematical relationship between the Gini--Simpson index and Shannon entropy is well-known \cite{Havrda1967, Keylock2005, Vajda2007} and has been demonstrated in the context of species diversity indices \cite{Zhang2014}.
But the connection has not been reported for the corresponding measures of nodal connection diversity in networks, $\mathcal{P}$ and $\mathcal{E}$.
The relationship can be seen through the series expansion of participation entropy via the logarithm in Eq.~\eqref{eq:participation-entropy}:
\begin{equation}
\label{eqn:series_expansion}
    \mathcal{E}_i(\mathcal{M}) = -\sum_{m\in\mathcal{M}} p_{i}(m) \sum_{n=1}^\infty \frac{-[1 - p_{i}(m)]^{n}}{n}\,.
\end{equation}
This quantity converges for $0 < p_{i}(m) \leq 1$, and we take $0 \log 0 \rightarrow 0$ by convention, so there is no contribution from any $p_{i}(m) = 0$.
Limiting the expansion to the leading term, $n = 1$, yields
\begin{align}
    \mathcal{E}_i(\mathcal{M}) & \approx \sum_{m\in\mathcal{M}} p_i(m) - p_{i}(m)^2\,, \nonumber \\
            &= 1 - \sum_{m\in\mathcal{M}} p_{i}(m)^2\,,\\
            &= \mathcal{P}_i(\mathcal{M})\,.
\end{align}
We thus recapitulate the participation coefficient as a first-order approximation to participation entropy (as per the Gini--Simpson index and Shannon entropy \cite{Zhang2014}).

\begin{figure}[h]
    \centering
    \includegraphics[width=.9\columnwidth]{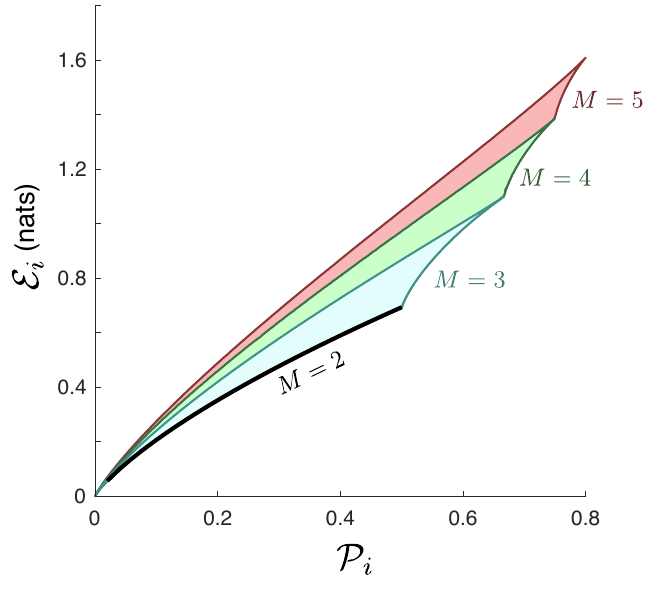}
    \caption{
    \textbf{Constraints on the relationship between $\mathcal{P}_i$ and $\mathcal{E}_i$ for networks containing $M = 2, ..., 5$ modules.}
    The shaded region for each $M$ indicates the additional allowed region, in addition to that accessible for lower $M$ values.
    }
    \label{fig:multipleModulesRegions}
\end{figure}

In order to investigate the discrepancy between $\mathcal{E}_i$ and its first-order approximation, $\mathcal{P}_i$, we sampled from possible distributions, $p_i(m)$ for $M = 2, ..., 5$, and plotted the resulting accessible regions of $\mathcal{P}_i$--$\mathcal{E}_i$ space in Fig.~\ref{fig:multipleModulesRegions}.
Our numerical results match analytic expressions for these regions for the underlying measures on $p_i(m)$ derived by \citet{Vajda2007}.
We find that $\mathcal{P}_i$ varies monotonically with $\mathcal{E}_i$ for $M = 2$, but for $M > 2$, allowed values of $\mathcal{P}_i$ and $\mathcal{E}_i$ are constrained to specific regions of the space.
This accessible region expands with the addition of each new module; Fig.~\ref{fig:multipleModulesRegions} annotates the additional accessible region with each increment of $M$.
The results indicate that there can be a substantial discrepancy between an analysis using $\mathcal{P}_i$ versus $\mathcal{E}_i$, with greater potential for differences at moderate-to-high values of $\mathcal{P}_i$ and with increasing $M$.
Published results using $\mathcal{P}_i$ to quantify nodal diversity (or extract a list of `high-participation nodes' \cite{Sporns2007, Pedersen2020}), may thus obtain different results when using $\mathcal{E}_i$ instead of its first-order approximation, $\mathcal{P}_i$.

\subsection*{Joint and conditional participation entropy}

A major advantage of formulating $\mathcal{E}_i$ as an entropy is the ability to capture more subtle types of connection-pattern diversity in networks.
Here we demonstrate this capability by developing entropy-based network participation measures for the case that each node is annotated with \textit{multiple} labels.
Specifically, we consider $L$ different module sets, $\mathcal{M}_1, \mathcal{M}_2, ..., \mathcal{M}_L$, that each define a labeling of network nodes.
In a social network, this could correspond to individuals being labeled by both gender, $\mathcal{M}_g$, and friendship group, $\mathcal{M}_f$.
Or, in a brain network, it could correspond brain regions being labeled by both their hemisphere, $\mathcal{M}_h$ (left or right) and their functional network module, $\mathcal{M}_f$ (e.g., auditory, visual, association, etc.).
There is no clear way of extending $\mathcal{P}_i$ to such a setting, but it can be incorporated naturally in the information-theoretic formulation of $\mathcal{E}_i$.

Extending participation entropy with respect to any single labeling of nodes, $\mathcal{M}$, we now consider the diversity of connections involving node $i$ across multiple label sets jointly.
Writing the $L$ sets as $\underline{\mathcal{M}} = (\mathcal{M}_1, \mathcal{M}_2, ..., \mathcal{M}_L)$, and a combination of labels from $\underline{\mathcal{M}}$ for a given node as $\underline{m} = (m^{(1)}, m^{(2)}, ..., m^{(L)})$, we define the joint probability distribution $p_i(\underline{m})$ for the connected neighbors of node $i$.
We can then define the \textit{joint participation entropy} of node $i$ as:
\begin{align} \label{eq:joint-participation-entropy}
    \mathcal{E}_i(\underline{\mathcal{M}}) & = H[p_i(\underline{m})] \nonumber \\
     & = - \sum_{\underline{m}} p_{i}(\underline{m}) \log p_{i}(\underline{m})\,.
\end{align}
This tells us the total diversity of connections across these multiple module sets, $\underline{\mathcal{M}}$.

Similarly, we can define the \textit{conditional participation entropy} as the entropy of modular assignments $\underline{m}$ from sets $\underline{\mathcal{M}}$ of the connected neighbors of node $i$, given knowledge of the modular assignments $\underline{n}$ from other sets $\underline{\mathcal{N}}$:
\begin{align}
    \mathcal{E}_i(\underline{\mathcal{M}} | \underline{\mathcal{N}}) & = H[p_i(\underline{m} | \underline{n})]\,, \nonumber \\
     & = \mathcal{E}_i(\underline{\mathcal{M}}, \underline{\mathcal{N}}) - \mathcal{E}_i(\underline{\mathcal{N}})\,.
\label{eq:chainRulePE}
\end{align}
This quantifies the remaining uncertainty in the distributions of connections across the modules of sets $\underline{\mathcal{M}}$, given that we already know their distributions across sets $\underline{\mathcal{N}}$.

The joint participation entropy, $\mathcal{E}_i(\underline{\mathcal{M}})$, and conditional participation entropy, $\mathcal{E}_i(\underline{\mathcal{M}} | \underline{\mathcal{N}})$, are related via the chain rule for entropies \cite{Cover2005} (vis-\`a-vis \eq{chainRulePE}), which means that we can consistently decompose and re-compose the diversity of connections over multiple module sets, regardless of which order we chain our knowledge of the module labelings.
This property is unique to the information-theoretic formulation \cite{Ash1965}.

\begin{figure}[h]
    \centering
    \includegraphics[width=0.48\textwidth]{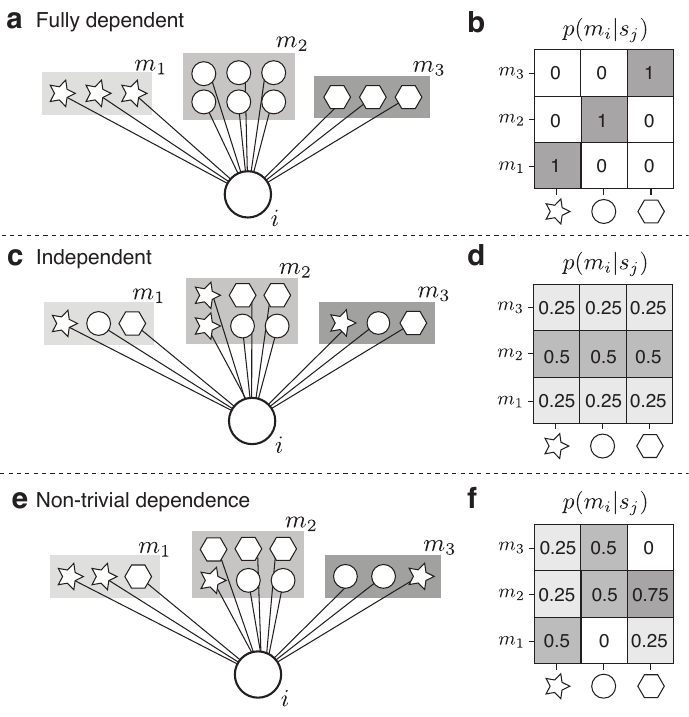}
    \caption{
    \textbf{Participation entropy can be extended to the case where each node is labeled according to multiple module sets.}
    Here we illustrate the case in which each node is labeled by two different module sets: $\mathcal{M} = \{m_1,m_2,m_3\}$ and $\mathcal{S} = \{$\FiveStarOpen,$\bigcirc$,$\hexagon$\}.
    Three cases are illustrated:
    \textbf{a}, \textbf{b}, labelings $\mathcal{M}$ and $\mathcal{S}$ are redundant;
    \textbf{c}, \textbf{d}, labelings $\mathcal{M}$ and $\mathcal{S}$ are statistically independent; and
    \textbf{e}, \textbf{f}, labelings $\mathcal{M}$ and $\mathcal{S}$ exhibit non-trivial dependence.
    \textbf{a}, \textbf{c}, \textbf{e} show the pattern of connectivity from a target node, $i$, to a set of nodes labeled by $\mathcal{M}$ and $\mathcal{S}$.
    \textbf{b}, \textbf{d}, \textbf{f} show conditional probability matrices, $p(m_i|s_j)$ for connected neighbors of node $i$ from both labelings.
    }
    \label{fig:schematic_conditional}
\end{figure}

To illustrate the calculation of conditional participation entropy, we show some illustrative examples in \fig{schematic_conditional} for the simple case of two node labelings: $\mathcal{M} = \{m_1, m_2, m_3\}$ and $\mathcal{S} = \{$\FiveStarOpen, $\bigcirc$, $\hexagon$\}.
The three cases shown in Fig.~\ref{fig:schematic_conditional} correspond to distinct types of connection patterns of node $i$ with respect to $\mathcal{M}$ and $\mathcal{S}$.
In \fig{schematic_conditional}a, the labels assigned to node $i$'s connected neighbors are \textit{redundant} with respect to $\mathcal{M}$ and $\mathcal{S}$.
That is, for a given connected neighbor, knowledge of the label $s$ leaves no uncertainty about the label $m$ (and vice-versa), resulting in the symmetric $p(m_i|s_j)$ matrix shown in \fig{schematic_conditional}b.
For this case, the conditional participation entropy of node $i$, $\mathcal{E}_i(\mathcal{M} | \mathcal{S}) = 0$.

For the connection pattern shown in \fig{schematic_conditional}c, the labelings $m$ and $s$ are statistically independent.
That is, for a given connected neighbor, knowledge of the label $s$ does not reduce our uncertainty about the label $m$, as reflected in the $p(m_i|s_j)$ matrix in \fig{schematic_conditional}d.
In this case, $\mathcal{E}_i(\mathcal{M} | \mathcal{S}) = \mathcal{E}_i(\mathcal{M})$.

In general, a node's connection pattern will involve non-trivial statistical dependencies between the combinations of labels.
Such a case is shown in \fig{schematic_conditional}e, where knowledge of the label $s$ reduces our uncertainty about $m$.
For example, as depicted in \fig{schematic_conditional}f, if we learn that a node is labeled $s = \bigcirc$, then our uncertainty about its label, $m$, is reduced, from $\{p(m_1),p(m_2),p(m_3)\} = \{0.25, 0.5, 0.25\}$ to $\{0, 0.5, 0.5\}$.
As such, $0 < \mathcal{E}_i(\mathcal{M} | \mathcal{S}) < \mathcal{E}_i(\mathcal{M})$ here.

The conditional participation entropy thus provides a new way to quantify a node's connection diversity across multiple labelings of network nodes.
For example, in a structural brain network in which brain areas (nodes) are annotated by both by a functional annotation, $\mathcal{M}_f$ (e.g., visual, auditory, motor, etc.) and their hemisphere, $\mathcal{M}_h$ (left or right), $\mathcal{E}(\mathcal{M}_h|\mathcal{M}_f)$ could be used to highlight nodes whose diversity of connectivity between left and right hemispheres depends on which functional module they connect to.


\subsection*{Conclusion}

We have introduced an information-theoretic formulation of nodal connection diversity in complex networks, incorporating results from the broader literature on quantitative diversity indices that builds on a prior introduction of the Shannon entropy formulation of participation coefficient \cite{RubinovSporns}.
Quantifying connection diversity as the average uncertainty in the module label of a connected neighboring node, termed participation entropy, $\mathcal{E}_i$, has mathematically favourable properties over the more commonly used participation coefficient.
Using a probabilistic formulation of the two measures, we show that the participation coefficient is a first-order approximation to the participation entropy (as per the relationship of the underlying measures of diversity \cite{Zhang2014}).
Using the additivity of participation entropy with respect to chaining probability distributions for multiple module sets, we introduce new ways of measuring connection diversity for cases where nodes are labeled from multiple label sets, defining joint and conditional participation entropy.

Future work may build on the theoretical foundations laid here, including applying the new measures to data.
This will require developing statistical significance tests against appropriate null distributions.
For example, analysis on the conditional participation entropy of a node, $\mathcal{E}_i(\mathcal{M} | \mathcal{N})$ (i.e., the diversity of connectivity across modules $\mathcal{M}$ given the labeling $\mathcal{N}$) requires comparison to an appropriate null hypothesis.
One choice of null hypothesis is that node $i$ connects randomly with respect to $\mathcal{M}$, while preserving the distribution of connections over $\mathcal{N}$ (which could be sampled from numerically).
Future work could also explore alternative probabilistic formulations of connection diversity that may differently account for module size \cite{Pedersen2020, RuizVargas2014:GatewayCoefficientNovel}.
In summary, the new theory introduced here enables practical new ways of understanding and quantifying more subtle types of nodal connection patterns in complex networks.



\bibliography{references}

\end{document}